\documentclass[useAMS,usenatbib]{mn2e}
\usepackage{rotating}

 \usepackage{times}

\def\pcm3{{\rm\thinspace cm^{-3}}}

\def\contcaption{\@conttrue\SFB@caption\@captype}

\def\n_h{{\rm n_{H}}}

\def\NH1{{$N_{\rm HI}~$}}

\def\ga{{\rm\thinspace gauss}}

\def\approxlt{\mathrel{\hbox{\rlap{\lower .5ex \hbox {$\sim$}}
        \raise .15 ex \hbox{$<$}}}}
\def\approxgt{\mathrel{\hbox{\rlap{\lower .5ex \hbox {$\sim$}}
        \raise .15 ex \hbox{$>$}}}}

\def\la{\mathrel{\hbox{\rlap{\hbox{\lower4pt\hbox{$\sim$}}}\hbox{$<$}}}}
\def\ga{\mathrel{\hbox{\rlap{\hbox{\lower4pt\hbox{$\sim$}}}\hbox{$>$}}}}

\newbox\grsign \setbox\grsign=\hbox{$>$} \newdimen\grdimen
\grdimen=\ht\grsign
\newbox\simlessbox \newbox\simgreatbox \newbox\simpropbox
\setbox\simgreatbox=\hbox{\raise.5ex\hbox{$>$}\llap
     {\lower.5ex\hbox{$\sim$}}}\ht1=\grdimen\dp1=0pt
\setbox\simlessbox=\hbox{\raise.5ex\hbox{$<$}\llap
     {\lower.5ex\hbox{$\sim$}}}\ht2=\grdimen\dp2=0pt
\setbox\simpropbox=\hbox{\raise.5ex\hbox{$\propto$}\llap
     {\lower.5ex\hbox{$\sim$}}}\ht2=\grdimen\dp2=0pt
\def\simgreat{\mathrel{\copy\simgreatbox}}
\def\simless{\mathrel{\copy\simlessbox}}

\title[Near-IR spectroscopy of DA white dwarfs]{A near-IR spectroscopic search for very-low-mass cool companions to notable DA white dwarfs $\thanks{Based on observations collected at the European Southern Observatory, Chile. ESO No. 072.D-0362}$}

\author[P. D. Dobbie et al.]{P. D. Dobbie$^{1}$\thanks{E-mail:
pdd@star.le.ac.uk} M. R. Burleigh$^{1}$ A. J. Levan$^{1}$ M. A. Barstow$^{1}$ R. Napiwotzki$^{1}$ \newauthor J. B. Holberg$^{2}$ I. Hubeny$^{3}$ and S.B. Howell$^{4}$ \\
$^{1}$Department of Physics and Astronomy, University of Leicester, University Road, Leicester LE1 7RH, UK\\
$^{2}$Lunar and Planetary Laboratory, Gould-Simpson Building, University of Arizona, Tucson, AZ 85721, USA \\
$^{3}$National Optical Astronomy Observatory, Tucson, AZ 85726, USA\\
$^{4}$WIYN Observatory and NOAO, 950 N. Cherry Ave., Tucson, AZ 85726}
\begin{document}

\date{Accepted 1988 December 15. Received 1988 December 14; in original form 1988 October 11}

\pagerange{\pageref{firstpage}--\pageref{lastpage}} \pubyear{2002}

\maketitle

\label{firstpage}

\begin{abstract}

We have undertaken a detailed near-IR spectroscopic analysis of eight notable white dwarfs, predominantly
of southern declination. In each case the spectrum failed to reveal compelling evidence for the presence of
a spatially unresolved, cool, late-type companion. Therefore, we have placed an approximate limit on the spectral-type 
of a putative companion to each degenerate. From these limits we conclude that if GD659, GD50, GD71 or 
WD2359-434 possesses an unresolved companion then most probably it is substellar in nature 
(M$<0.072$M$_{\odot}$). Furthermore, any spatially unresolved late-type companion to RE J0457-280, 
RE J0623-374, RE J0723-274 or RE J2214-491 most likely has M$<0.082$M$_{\odot}$. These results
imply that if weak accretion from a nearby late-type companion is the cause of the 
unusual photospheric composition observed in a number of these degenerates then the companions are of very low mass, beyond the detection thresholds of this study. Furthermore, these results do 
not contradict a previously noted deficit of very-low-mass stellar and brown dwarf companions to main sequence
 F,G,K and early-M type primaries ($a\simless1000$\,AU). 

\end{abstract}


\section{Introduction}

Examples of white dwarfs in binary systems with early-mid M dwarf companions are plentiful. Many are resolved
as common proper motion pairs (e.g. Silvestri et al. 2001), while a number of unresolved systems have recently been 
identified through near-infrared photometry (e.g Green, Ali \& Napiwotzki 2000). A further proportion are revealed
in optical spectra either as excess red emission or through the detection of narrow emission components in the 
cores of the white dwarf's HI Balmer absorption lines (e.g. Thorstensen, Vennes \& Shambrook 1994). This emission 
may either be intrinsic to an active cool companion (e.g. RE~J1629$+$780, Cooke et al. 1992, Sion et al. 1995), or 
be due to the irradiation of the atmosphere of the cool star facing the hot white dwarf, particularly if the system 
components are close (e.g. Vennes \& Thorstensen 1994). The latter systems are of interest as one outcome of common 
envelope (CE) evolution and/or as the precursors of cataclysmic variable systems (pre-CVs). Some may even be old 
CVs during a period of no mass transfer (Howell \& Ciardi 2001). As relatively few of these close systems are
known, when discovered they are intensively studied spectroscopically to obtain mass functions from the radial 
velocity curves (e.g. Vennes, Thorstensen \& Polomski 1999).   

In contrast, very few white dwarfs with late-M and cooler companions of very-low stellar or substellar mass
are known (e.g Farihi et al. 2003). This is likely due, in part, to the spectral energy distribution of such 
systems being dominated at optical wavelengths by the white dwarf, particularly if the degenerate star is 
relatively hot (T$_{\rm eff}\ga10,000$K). In these systems the companion instead may be revealed 
photometrically as an infrared excess e.g. the white dwarf $+$ L4 dwarf pair GD165 (Becklin \& Zuckerman 1988,
Kirkpatrick et al. 1999) or through the detection of features characteristic of a cool dwarf in a near-IR 
spectrum. Indeed, Farihi \& Christopher (2004) have recently announced the detection from 2MASS photometry and 
K band IR spectroscopy of an unresolved companion with spectral type L5.5 or later to the ZZ Ceti white 
dwarf GD1400. This likely represents the first unambiguous detection of a substellar companion to a white dwarf. 
However, despite Makarov (2004) having recently claimed the existence of a substellar companion with a mass $0.06\pm0.02$M$_\odot$ to the nearest isolated white dwarf, van Maanen 2, on the basis of Hipparcos astrometric measurements, a 
subsequent attempt to directly detect this brown dwarf, in adaptive optics images obtained in the L band and 
archival mid-infrared ISO observations, has been unsuccessful. Therefore, his interpretation is now virtually ruled
out (Farihi, Becklin \& MacIntosh 2004). Further, although Wachter et al. (2003) have recently identified 47 new 
unresolved white dwarf $+$ red dwarf binaries from 2MASS photometry, they are unable to claim any as bona 
fide very-low-mass stars or brown dwarfs. 

\begin{table*}
\begin{minipage}{170mm}
\begin{center}
\caption{Summary details of the white dwarfs studied in this work, including near-IR magnitudes for 
each star obtained from the 2MASS All-Sky Point Source Catalogue. The exposure times used for
 the acquisition of the JH and HK near-IR spectra with the NTT and SOFI are also listed.}
\label{sum1}
\begin{tabular}{llllcccrr}
\hline
Identity   & Name & \multicolumn{1}{|c|}{RA}    &  \multicolumn{1}{|c|}{Dec}   & J &  H & K$_{\rm S}$ & \multicolumn{2}{|c|}{t$_{\rm exp}$ (secs)} \\
& & \multicolumn{2}{|c|}{J2000.0} &  &  & & \multicolumn{1}{|c|}{JH} &  \multicolumn{1}{|c|}{HK}  \\
 \hline
WD0050-332 & GD659        & 00 53 17.43  & -32 59 56.5 & $14.00\pm0.03$  & $14.17\pm0.05$ & $14.32\pm0.08$ &  720 & 2400  \\
WD0346-011 & GD50         & 03 48 50.20  & -00 58 31.2 & $14.75\pm0.03$  & $14.86\pm0.04$ & $15.12\pm0.14$ & 1080 & 2880  \\
WD0455-282 & RE J0457-280 & 04 57 13.9   & -28 07 54   & $14.68\pm0.03$  & $14.85\pm0.07$ & $14.72\pm0.11$ & 1080 & 2880  \\
WD0549+158 & GD71         & 05 52 27.62  & +15 53 13.3 & $13.73\pm0.03$  & $13.90\pm0.04$ & $14.12\pm0.07$ &  720 & 1920  \\
WD0621-376 & RE J0623-374 & 06 23 12.2   & -37 41 29   & $12.85\pm0.03$  & $12.96\pm0.02$ & $13.09\pm0.03$ &  360 &  960  \\
WD0721-276 & RE J0723-274 & 07 23 19.8   & -27 47 17   & $15.30\pm0.07$  & $15.37\pm0.14$ & $15.36\pm0.21$ & 1080 & 2880  \\
WD2211-495 & RE J2214-491 & 22 14 11.91  & -49 19 27.3 & $12.44\pm0.03$  & $12.61\pm0.03$ & $12.64\pm0.03$ &  360 &  960  \\ 
WD2359-434 & LHS 1005     & 00 02 10.75  & -43 09 55.6 & $12.60\pm0.03$  & $12.43\pm0.02$ & $12.45\pm0.02$ &  480 &  1440  \\
\hline
\end{tabular}
\end{center}
\end{minipage}
\end{table*}

Nevertheless, the detection of white dwarf + very-low mass stellar or brown dwarf binaries is 
an important issue. In some apparently isolated white dwarfs, there exist puzzling heavy element abundance 
anomalies which appear to imply some external source of accretion. This is because the timescales for the 
gravitational settling of heavy elements are very short compared to the white dwarf cooling time (e.g Dupuis 
et al. 1993). Unless this high-Z material is accreted from very dense interstellar clouds, there must exist
some unseen source associated with the white dwarf itself. For example, high levels of refractory elements 
such as Mg, Al, Si and Ca are observed in the photosphere of EG\,102 yet the residence time for these on the
white dwarf surface is only 3 days (Holberg et al. 1997). Further, unexpectedly large abundances of Ca, Mg 
and Fe are observed in the atmosphere of the massive cool hydrogen rich white dwarf GD362. Gianninas, Dufour 
\& Bergeron (2004) argue that it is unlikely the origin of these metals is accretion of interstellar material.
Indeed, Zuckerman et al. (2003) have recently noted the high frequency of DAZ stars which are part of known 
binary systems and have suggested that wind driven mass loss from the red dwarf companion may be responsible for
a proportion of the observed DAZ stars. 

Additionally, observations of such systems can be of use in placing limits on the fraction of normal stars 
with very-low-mass stellar or substellar companions at $a\simless1000$\,AU, when part of a statistically 
robust campaign. This range of separation is of particular interest as current radial velocity and 
coronographic imaging surveys may indicate a discrepancy between the brown dwarf companion fraction at small
separation ($a<1000$\,AU; $1\pm1$\%; McCarthy \& Zuckerman 2004, Marcy and Butler, 2000) and large 
radii ($a>1000$\,AU; $10-30$\%; Gizis et al. 2001), although the validity of this latter result is still 
debated. Indeed, models of binary evolution which include a population of zero-age main sequence 
+ substellar pairs predict the existence of a population of cataclysmic variables with orbital periods
$\simless2.5$hrs and which contain brown dwarf secondaries (e.g Politano 2004). Despite a small number 
of candidates for such systems having been found, e.g. Howell \& Ciardi (2000) believe they have directly 
detected via infrared spectroscopy  substellar secondaries in LL And and EF Eri, all lie above the observed
CV period minimum ($\sim75$mins) and evolutionary models suggest it more likely that the secondaries in
these systems have evolved through mass loss to become substellar as opposed to them being born brown dwarfs.

\begin{figure*}
\vspace{360pt}
\includegraphics{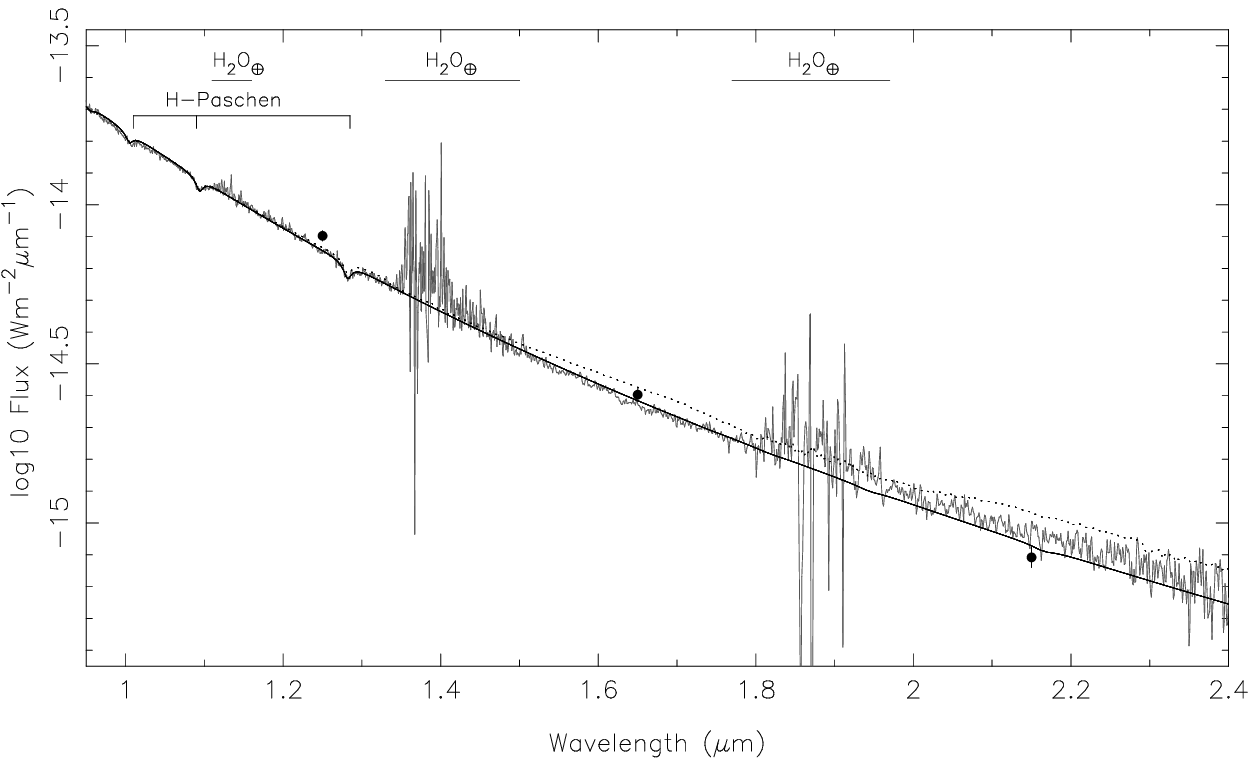}
\includegraphics{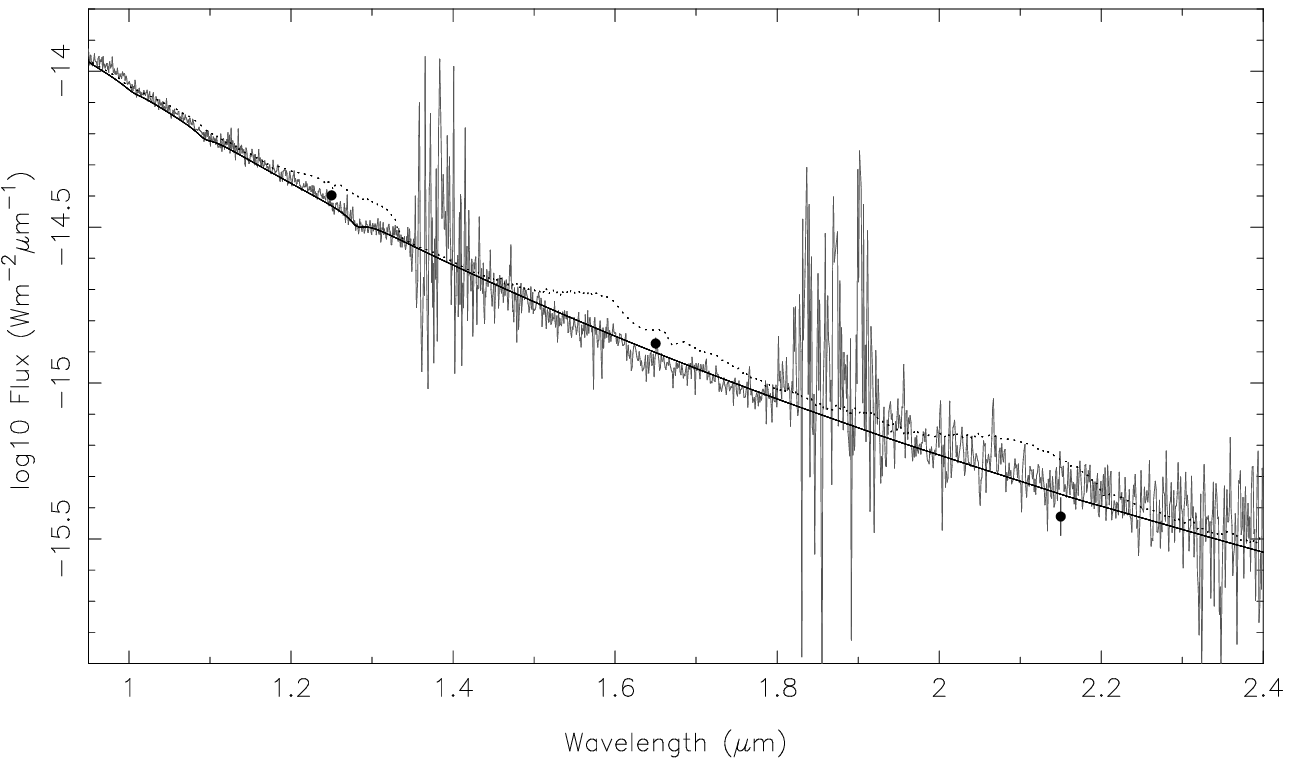}
\caption{a. Near-IR spectroscopy (solid grey lines) and 2MASS JHK photometry (filled circles) of the white dwarfs WD0050-332 and WD0346-011 in the top and bottom panels respectively. Pure-H non-LTE synthetic white dwarf spectra of appropriate effective temperature, surface gravity and normalisation (solid black lines) and hybrid white dwarf + late-type dwarf models representing our estimated limits on the spectral type of putative spatially unresolved companions (dotted black line: top, WD+L6 and bottom, WD+T5) are overplotted. In the top plot we have labelled the more prominent white dwarf HI Paschen and telluric water vapour line features present in these datasets.}
\end{figure*}

\setcounter{figure}{0}

\begin{figure*}
\vspace{360pt}
\includegraphics{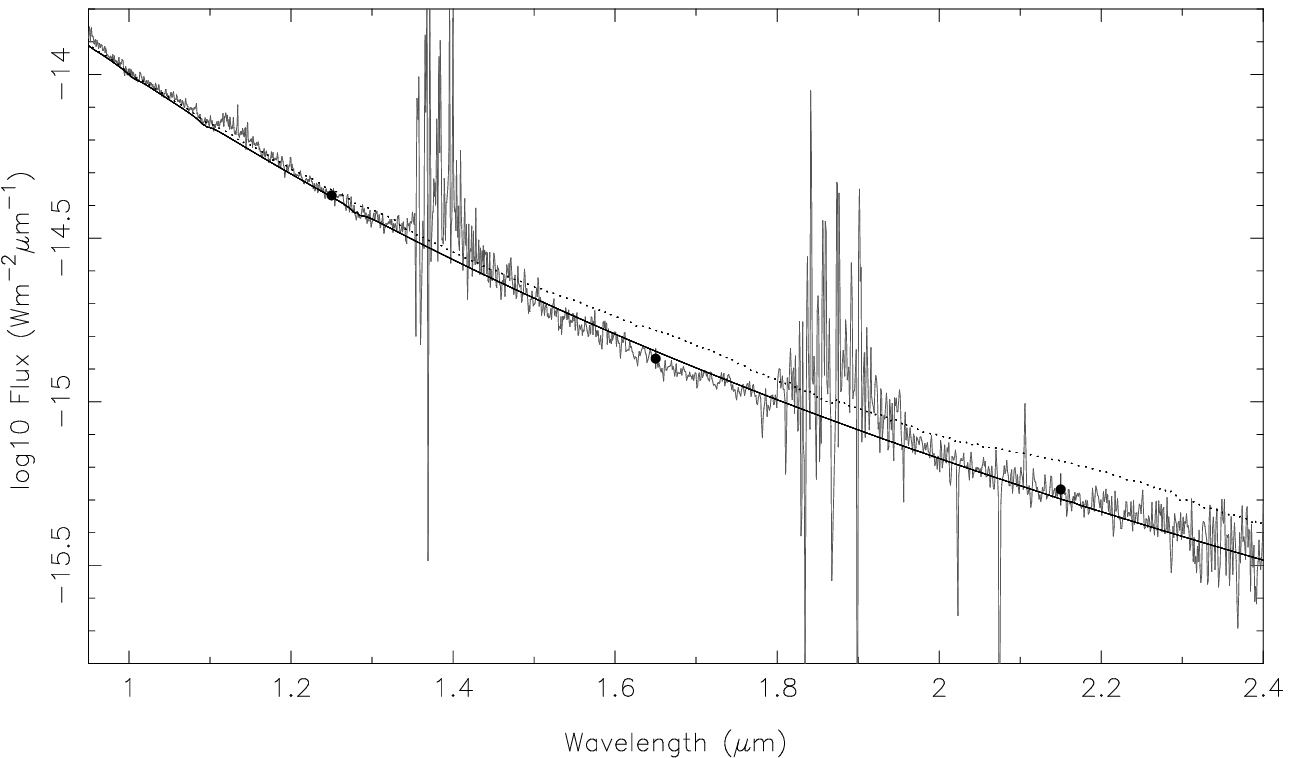}
\includegraphics{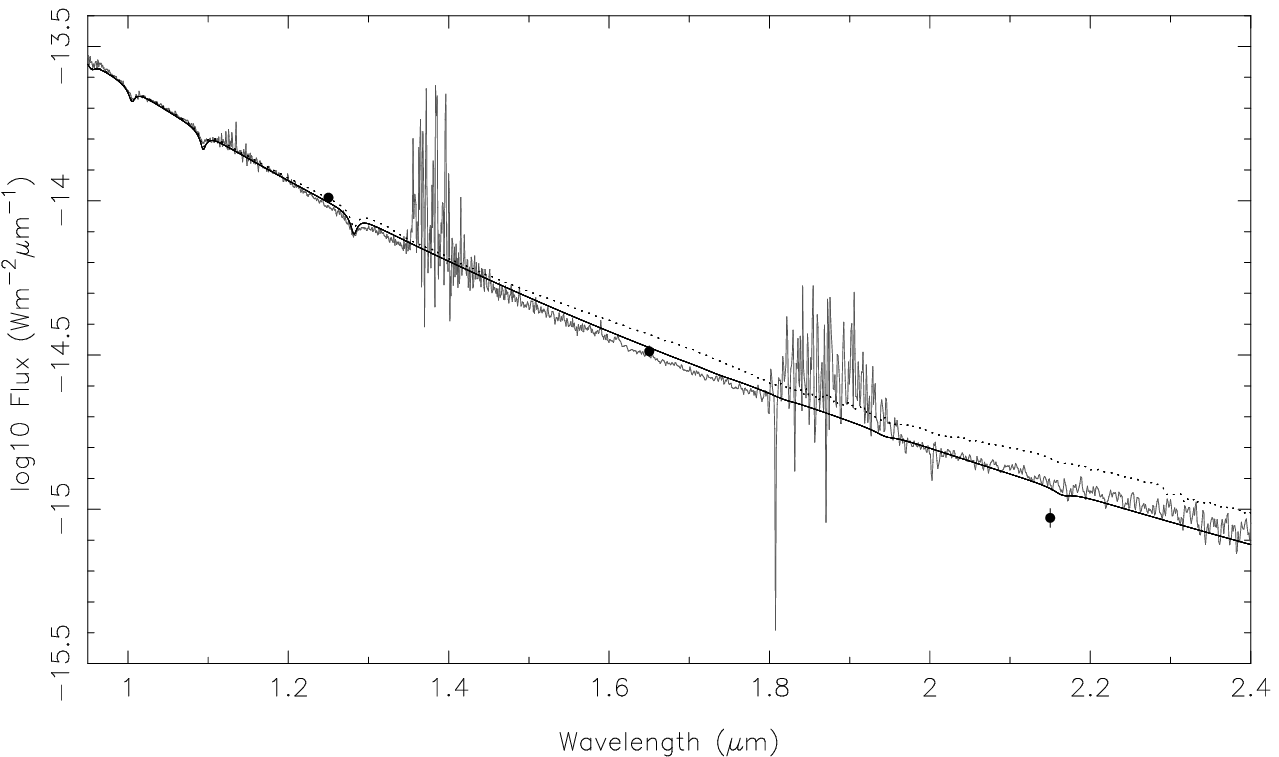}
\caption{b. top: WD0455-282 (WD+L3), bottom: WD0549+158 (WD+L6) - symbols as in Figure 1a.}
\end{figure*}

\setcounter{figure}{0}

\begin{figure*}
\vspace{360pt}
\includegraphics{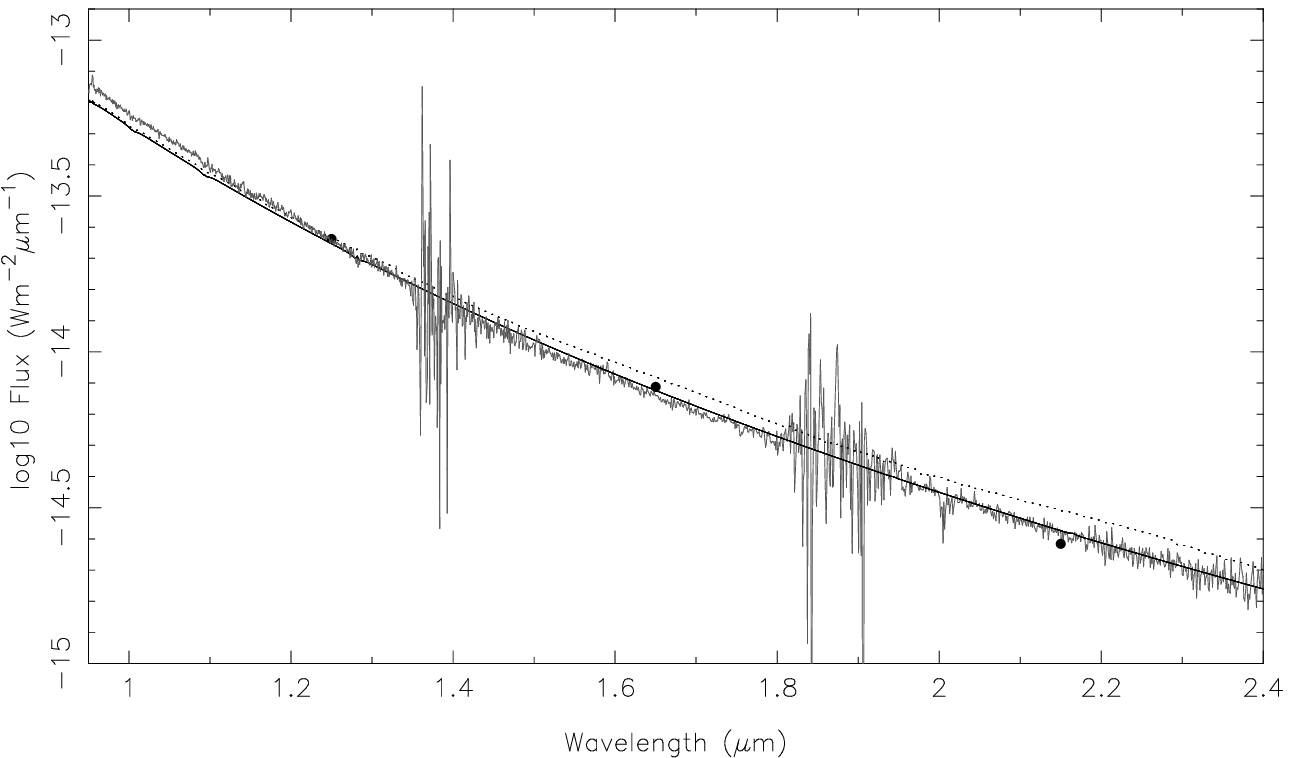}
\includegraphics{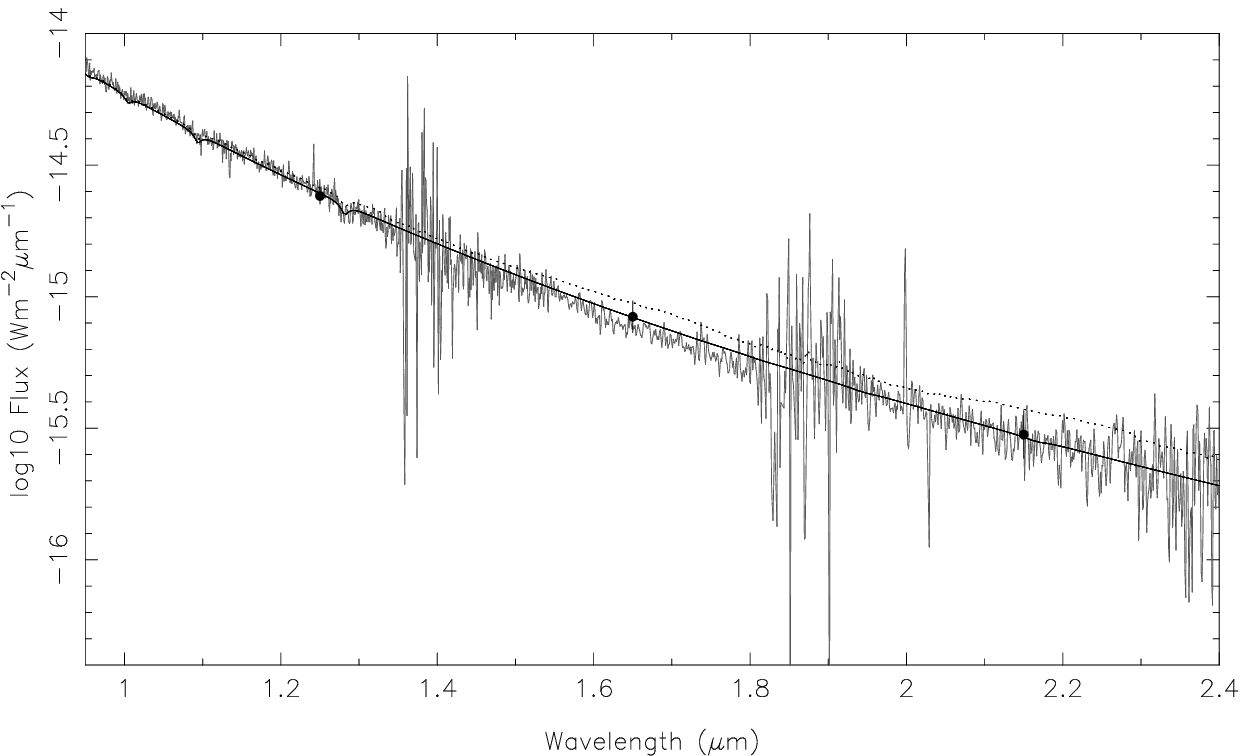}
\caption{c. top: WD0621-376 (WD+M9), bottom: WD0721-276 (WD+L4) - symbols as in Figure 1a.}
\end{figure*}

\begin{table*}
\begin{minipage}{170mm}
\begin{center}
\caption{Summary of the additional physical parameters for each white dwarf used in this work. For all objects 
excluding WD2359-434 and WD0721-276, effective temperatures, surface gravities and visual magnitudes have been
taken from the work of Marsh et al. (1997) and distances estimated using the evolutionary models of Wood (1995). 
While estimates of the effective temperature and surface gravity of WD0721-276 are also based on the work of
Marsh et al. (1997) the visual magnitude has been obtained from Wolff et al. (1999). The physical parameters we 
adopt for WD2359-434 have been provided by R. Napiwotzki; the distance of WD2359-434 has been derived from 
parallax measurements.}
\label{sum2}
\begin{tabular}{llllllc}
\hline

Identity   & Name & T$_{\rm eff}$(K) & log g & \multicolumn{1}{|c|}{V} & D(pc) & Refs \\
\hline
WD0050-332 & GD659        &  34684 & 7.89 &  $13.37\pm0.02$ &  57   & 1  \\
WD0346-011 & GD50         &  42373 & 9.00 &  $14.04\pm0.02$  &  33  & 1  \\
WD0455-282 & RE J0457-280 &  58080 & 7.90 &  $13.951\pm0.009$ & 104  & 1 \\
WD0549+158 & GD71         &  32008 & 7.70 &  $13.032\pm0.009$ & 51  & 1  \\
WD0621-376 & RE J0623-374 &  62280 & 7.22 &  $12.089\pm0.001$ & 82  & 1  \\
WD0721-276 & RE J0723-274 &  37120 & 7.75 &  $14.52\pm0.1$ & 113 & 1,2 \\
WD2211-495 & RE J2214-491 &  65600 & 7.42 &  $11.708\pm0.009$ & 60 & 1  \\
WD2359-434 & LHS 1005     &  8660  & 8.56 &  $13.05\pm0.02$ & 7.8 & 3 \\
\hline
\end{tabular}
\end{center}

1. Marsh et al. (1997) \\
2. Wolff et al. (1999) \\
3. Aznor Cuadrado et al. (2004) \\

\end{minipage}
\end{table*}

\setcounter{figure}{0}
\begin{figure*}
\vspace{360pt}
\includegraphics{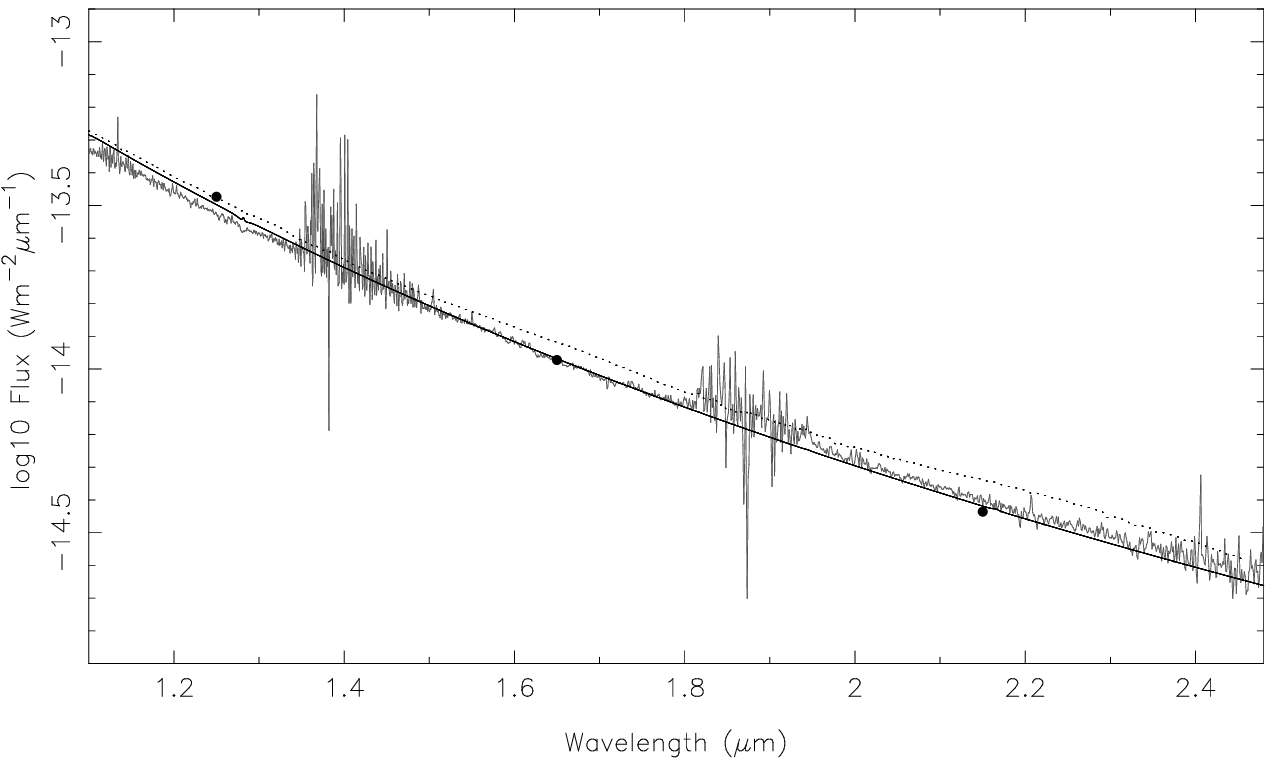}
\includegraphics{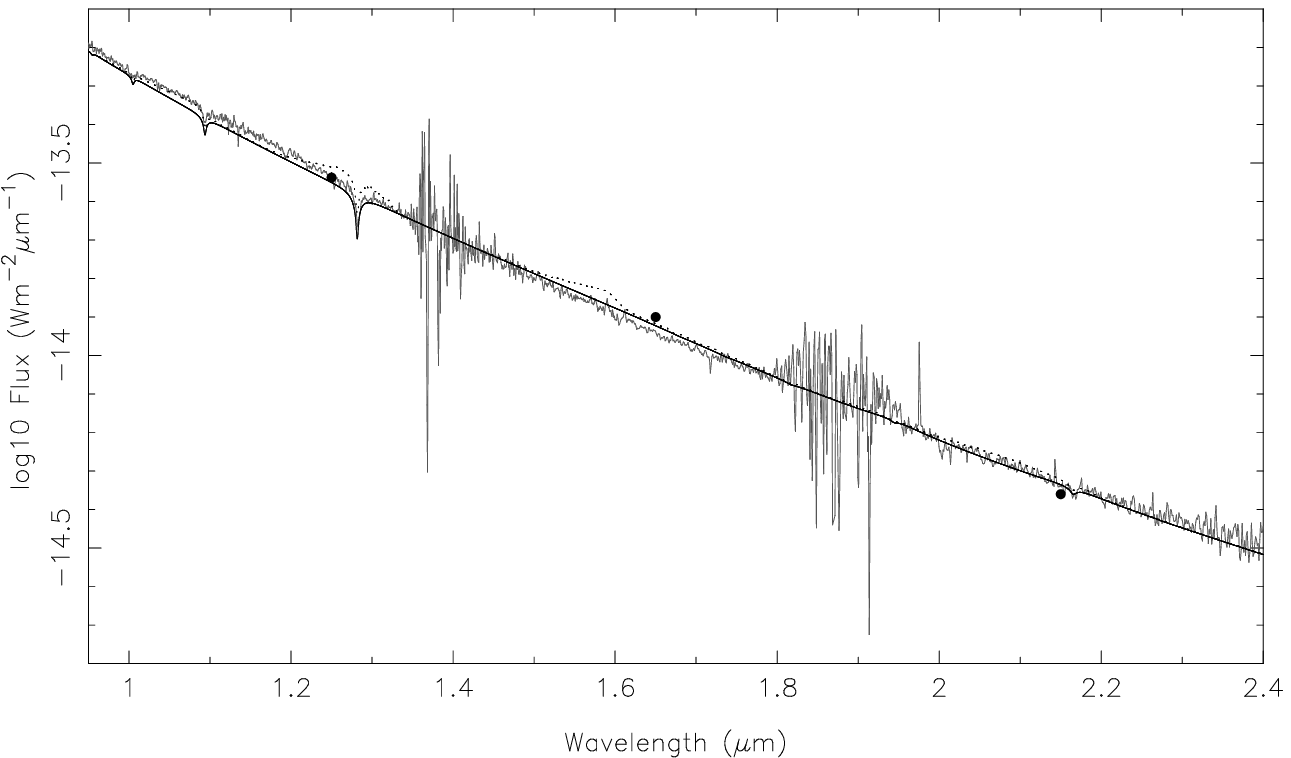}
\caption{d. top: WD2211-495 (WD+M9), bottom: WD2359-434 (LTE WD+T8) - symbols as in Figure 1a. Note in the bottom plot the features attributable to a putative T8 companion are comparable in size to the H-Paschen lines, which are clearly detected and match closely the model predictions.}
\end{figure*}

We have recently instigated an observational programme to obtain near-IR spectroscopy of notable 
DA white dwarfs the aim of which is to search for very-low-mass (mid-late-M, L and possibly T dwarf) 
companions which may explain e.g. peculiarities in their measured photospheric compositions or in their 
observed energy distributions. The presence of low-mass companions may be revealed through a subtle excess in 
continuum emission above that expected from the white dwarf alone, as well as through absorption 
signatures typical to cool dwarfs such as NaI, KI and H$_{2}$O. Our targets are predominantly 
hot objects (T$_{\rm eff}\simgreat30000$K) which have been studied extensively at EUV, FUV, UV and 
optical wavelengths but for which with no detailed examination at near-IR wavelengths has been published. 
More specifically we have concentrated on objects with no robust detection of a companion in existing
datasets but 1) with unexplained abundance anomalies, 2) with significant residuals between their observed 
Balmer lines and synthetic profiles or narrow HI Balmer emission lines, 3) which show radial velocity variations
or 4) well known white dwarf standards. Here we report the results based on our 
observations conducted with the NTT for a small collection of such degenerates with predominantly southern
declinations. 
 
\section[]{Near-IR spectrscopy}
\subsection{Observations}
Low resolution near-IR spectra were obtained for a number of white dwarfs with predominantly southern declinations using the ESO New Technology Telescope (NTT) and the Son-of-Isaac (SOFI) infrared instrument on 2003/12/09 and 2003/12/10. The sky conditions at the La Silla site were good on both nights with seeing typically in the range 0.6''-1.0'' with some small patches of cirrus cloud confined to the region of sky close to the southern horizon. SOFI operates at the Naysmyth A focus of the NTT and includes a Rockwell Hg:Cd:Te Hawaii detector with 1024x1024 18.5$\mu$m pixels. In the low resolution spectroscopic mode ($\lambda/\delta\lambda\sim950$ with the 0.6'' slit), as used for this work, coverage of the wavelength ranges $0.95-1.64\mu$m and $1.53-2.52\mu$m is provided by the ``blue'' and the ``red'' grism respectively. The observations were undertaken using the standard technique of nodding our point source targets back and forth along the spectrograph slit in an ABBA pattern. The individual on target exposure times were chosen to ensure that while the data were background limited, the sky was sampled frequently and the total counts in each pixel were comfortably within the linear regime of the detector ($\simless10000$ ADU). To minimise detector overheads we used the double correlated read mode which is well suited to these low resolution spectroscopic observations. The total integration times used for each white dwarf with the blue and the red grism are shown in Table~\ref{sum1}. To facilitate the removal of telluric features from the target spectra and to provide an approximate flux calibration, a standard star was observed either immediately before or after each science integration. These were carefully chosen to lie within $\sim0.1$ airmasses of each white dwarf. In addition, regular exposures of the xenon lamp were obtained to permit the reliable wavelength calibration of the spectra.

\subsection{Data reduction}

To reduce the data we have applied standard techniques using software routines in the STARLINK packages KAPPA and FIGARO. In brief, a bad pixel mask was constructed by merging a list of anomolously valued pixels clipped from dark frames with a generic map of bad array elements obtained from the ESO SOFI webpages.\footnote{www.ls.eso.org/lasilla/sciops/ntt/sofi/index.html} This was applied to all the data. The science, standard star and arc lamp spectral images were flat fielded with a normalised response map appropriate to either the blue or the red grism setup. Subsequently, difference pairs were assembled from the science and standard star images and any significant remaining sky background removed by subtracting linear functions, fitted in the spatial direction, from the data. The spectra of the white dwarfs and the standard stars were then extracted and assigned the wavelength solution derived from the relevant arc spectrum. Any features intrinsic to the energy distributions of the standard stars were identified by reference to a near-IR spectral atlas of fundamental MK standards (Wallace et al. 2000, Meyer et al. 1998, Wallace \& Hinkle 1997) and were removed by linearly interpolating over them. The spectrum of each white dwarf was then co-aligned with the spectrum of its standard star by cross-correlating the telluric features present in the data. The science spectra were divided by the standard star spectra and multiplied by a blackbody with the standard star T$_{\rm eff}$, taking into account the differences in exposure times. Finally, the flux levels were scaled to (1) achieve the best possible agreement between the blue and the red spectrum of each white dwarf in the overlap region between $1.53-1.64\mu$m and (2) obtain the best possible agreement between the spectral data and the J, H and K$_{\rm S}$ photometric fluxes for each object derived from the 2MASS All Sky Data Release Point Source Catalogue magnitudes (Skrutskie et al. 1995) where zero magnitude fluxes were taken from Zombeck (1990). The reduced spectra and 2MASS fluxes are shown in Figures 1a-d. 

\section{Analysis of the data} 

\subsection{Model white dwarf spectra}

For each object in our collection we have generated a pure-H synthetic white dwarf spectrum at the effective temperature and surface gravity given in Table~\ref{sum2}. We have used the latest versions of the plane-parallel, hydrostatic, non-local thermodynamic equilibrium (non-LTE) atmosphere and spectral synthesis codes TLUSTY (v200; Hubeny 1988, Hubeny \& Lanz 1995) and SYNSPEC (v48; Hubeny, I. and Lanz, T. 2001, ftp:/tlusty.gsfc.nasa.gov/synsplib/synspec). All calculations included a full treatment of line blanketing and used a state-of-the-art model H atom incorporating the 8 lowest energy levels and one superlevel extending from n=9 to n=80, where the dissolution of the high lying levels was treated by means of the occupation probability formalism of Hummer \& Mihalas (1988), generalised to the non-LTE situation by Hubeny, Hummer \& Lanz (1994). During the calculation of the model structure the lines of the Lyman and Balmer series were treated by means of an approximate Stark profile (Hubeny et al. 1994) but in the spectral synthesis step detailed profiles for these and the Paschen and Brackett lines were calculated from the Stark broadening tables of Lemke (1997). The synthetic spectral fluxes have been normalised to the V magnitude of the relevant white dwarf (Table~\ref{sum2}) and convolved with a Gaussian to match the resolution of the SOFI spectra. These are shown overplotted on the observed data in Figures 1a-d. It is worth noting here that in the effective temperature regime spanned by most of these white dwarfs, the colours V-K, J-H and H-K are rather weak functions of T$_{\rm eff}$ (e.g Bergeron, Wesemael \& Beauchamp 1995). 

\subsection{Searching for cool companions}

We have examined in turn each of the plots, Figures 1a-d, top and bottom, for significant differences in the overall shape or level between the observed and synthetic fluxes which can be consistent with the presence of a cool companion. Additionally, we have searched for specific features in each spectrum typical of the energy distributions of M, L or T dwarfs e.g. K~I and Na~I absorption at 1.25$\mu$m and 2.20$\mu$m respectively, CH$_{4}$ or CO at 1.6$\mu$m and 2.3$\mu$m respectively and H$_{2}$O centred on 1.15, 1.4 and 1.9$\mu$m. When no convincing evidence for such has been found, we have instead added empirical models for low mass stellar or substellar objects to the white dwarf synthetic spectrum and compared these composites to the IR data to obtain approximate limits on the spectral type of putative cool companions. The empirical models have been constructed using the near-IR spectra of M, L and T dwarfs presented by McLean et al. (2003). In brief, the data have been obtained with the NIRSPEC instrument on the Keck Telescope, cover the range $0.95-2.31\mu$m with a resolution of $\lambda/\delta\lambda\sim2000$ and have been flux calibrated using J, H and K$_{\rm S}$ photometric fluxes derived from the 2MASS magnitudes as described by McLean et al. (2001). To extend these data out to $2.4\mu$m, our effective red limit, we have appended to them sections of CGS4 spectra of late-type dwarfs obtained by Leggett et al. (2001) and Geballe et al. (2002). To match the resolutions of the NIRSPEC and SOFI spectra we have convolved the former with a Gaussian. The smaller difference in resolution between the SOFI and CGS4 data ($\lambda/\delta\lambda\sim600-900$) has been neglected. 

The fluxes of the empirical models have been scaled to a level appropriate to a location at d=10pc using the 2MASS J magnitude of each late-type object and the polynomial fits of Dahn et al. (2002) and Tinney et al. (2003) to the M$_{\rm J}$ versus spectral type for M6-M9 and L0-T8 field dwarfs/brown dwarfs respectively. Subsequently, these fluxes have been re-calibrated to be consistent with the distance of each white dwarf as derived from measured V magnitude and effective temperature and theoretical M$_{\rm V}$ and radius from evolutionary models of pure-C core white dwarfs with He and H layer masses of $10^{-2}$M$_{\odot}$ and $10^{-4}$M$_{\odot}$ respectively (Wood 1995). Further, in setting limits on companion spectral type, the fluxes of the empirical models have been reduced by a factor 1.4, corresponding to the rms dispersion in the M$_{\rm J}$ versus spectral type relationship of Tinney et al. (2003). Starting with T8 we have progressively added earlier spectral types to the synthetic white dwarf spectrum, until it could be concluded with reasonable certainty that the presence of a companion of that effective temperature or greater would have been obvious from our data, given the S/N. Propogating the typical errors in V, T$_{\rm eff}$ and log g, we find that the uncertainties in the estimated white dwarf distances due to measurement errors are of the order 5-6\% and have little impact on the limits we set on companion spectral type. We note also that the systematic errors which may be present in the effective temperatures we have adopted for the hotter white dwarfs in our study ($\sim10$\%), arising due to the LTE nature (as opposed to non-LTE) of the models used by Marsh et al. (1997) and their neglect of metal line blanketing, are in such a way that the limits we place err on the side of conservatism, at least for these white dwarfs. 

\begin{figure*}
\vspace{200pt}
\includegraphics{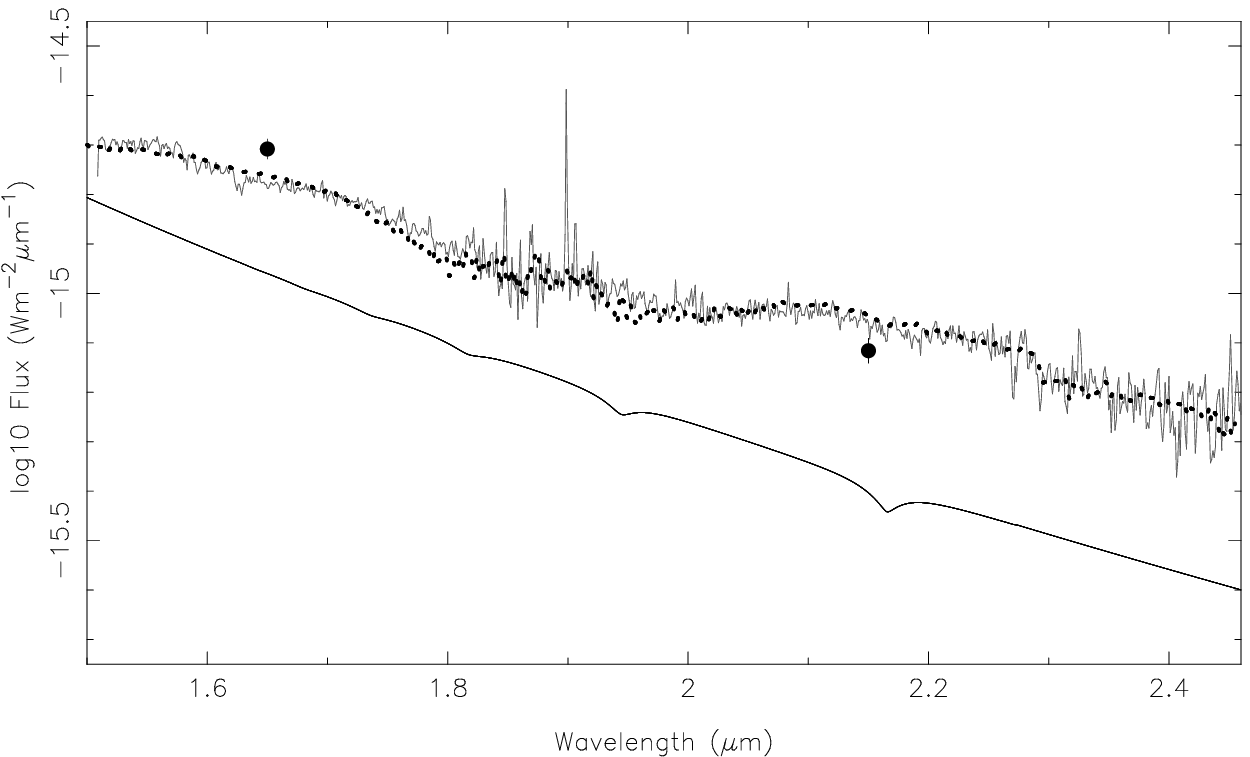}
\caption{Near-IR spectroscopy (solid grey line) and 2MASS H and K photometry (filled circles) of the white dwarf GD1400. A pure-H LTE synthetic white dwarf spectrum of appropriate effective temperature, surface gravity and normalisation (solid black line) and a hybrid white dwarf + late-type dwarf model representing our best estimate of the spectral type of the spatially unresolved companion (dotted black line - LTE WD+L7) are overplotted.}
\end{figure*}

To perform some means of assessment of our spectroscopic calibration and modelling we have recently obtained,
using the same telescope and instrument setup, a HK spectrum of the newly identified DA WD+dL binary 
GD1400 (Farihi \& Christopher 2004). These authors have concluded that the 2MASS photometry of this 
system is most consistent with a companion spectral type of L6. Following a similar procedure to that 
outlined above, using 2MASS H and K$_{\rm S}$ photometry to calibrate the NTT data and adopting the parameters 
for the white dwarf given in Farihi \& Christopher (2004), we have compared various combinations of 
synthetic white dwarf spectrum + low-mass stellar or substellar model to the observed data. We find 
that the best match to the data is provided by a WD+L7 model (see Figure 2) in satisfying  
agreement with Farihi \& Christopher's estimated spectral type given the uncertainties in their 
deconvolved near-IR photometry. The excellent level of agreement between the shapes of the composite model 
and the observed spectrum (better than 10\% where S/N allows) suggest that our data reduction and calibration 
procedures are reasonably robust and supports our use of the NIRSPEC datasets as low mass stellar and substellar
templates. 

\section{Results and Discussion}

\setcounter{table}{1}
\begin{table*}
\begin{minipage}{135mm}
\begin{center}
\caption{Limiting spectral types, temperatures and masses of cool companions to the white dwarfs in our collection.
 The approximate effective temperature of an object of this spectral type, as estimated from the polynomial relation detailed in Table 4 of Golimowski et al. (2004), is also shown. Furthermore, we provide rough upper limits on the masses as a function of age of putative cool companions, by comparing these effective temperatures to the predictions of 
the evolutionary models of Baraffe et al. (2003).   
 }
\label{tab2}
\begin{tabular}{llcrrrr}
\hline
ID & Name & SpT  & T$_{\rm eff}$(K) & 1Gyr(M$_{\odot}$) & 5Gyr(M$_{\odot}$) & 10Gyr(M$_{\odot}$) \\
\hline
WD0050-332 & GD659        & L6 & 1600  & 0.052 & 0.071  & 0.072   \\ 
WD0346-011 & GD50         & T5 & 1200  & 0.037 & 0.063  & 0.069   \\
WD0455-282 & RE J0457-280 & L3 & 1950  & 0.065 & 0.074  & 0.075   \\
WD0549+158 & GD71         & L6 & 1600  & 0.052 & 0.071  & 0.072   \\
WD0621-376 & RE J0623-374 & M9 & 2400  & 0.081 & 0.082  & 0.082   \\
WD0721-276 & RE J0723-274 & L4 & 1800  & 0.060 & 0.073  & 0.073   \\
WD2211-495 & RE J2214-491 & M9 & 2400  & 0.081 & 0.082  & 0.082   \\
WD2359-434 & LHS 1005     & T8 &  750  & 0.020 & 0.039  & 0.048   \\
\hline
\end{tabular}
\end{center}
\end{minipage}
\end{table*}

\subsection{Do we detect unresolved late-type companions to these white dwarfs $?$}
 
Green, Ali \& Napiwotzki (2000) searched J and K band photometry of 47 extreme-ultraviolet selected degenerates,
drawn from the catalogues of the EUVE All-Sky and ROSAT Wide Field Camera surveys, for a $3\sigma$ excess in 
both bands with respect to the predictions of white dwarf models. This led to the identification of 10 marginally 
resolved or unresolved white dwarf + dM systems, half of which were not previously suspected of being composite
in nature. A more recent analysis of several hundred white dwarfs drawn from the McCook \& Sion (1999) catalogue 
and present in the 2MASS Second Incremental Point Source Catalogue, has revealed, on the basis of their location 
in the J-H, H-K colour-colour diagram, 95 candidate white dwarf + red dwarf binaries, 47 of which were previously
unknown (Wachter et al. 2003). We note that GD50 and RE J0457-280 were included in both these photometric surveys 
and GD659 in the Wachter et al. (2003) study but none of the three were flagged as a likely unresolved white dwarf 
+ red dwarf composite.
 
Nevertheless, the present spectroscopic investigation allows us to probe to cooler spectral types and hence 
slightly lower masses. For example, by demanding a $3\sigma$ flux excess at J in addition to K, Green, Ali \& Napiwotzki (2000) 
effectively limit their search to companions with types earlier than late-M. Furthermore, using synthetic 2MASS 
colours generated from a number of white dwarf and composite white dwarf + red dwarf models we have examined the
location in the J-H, H-K colour-colour diagram of various combinations and find that for white dwarfs with effective temperatures and surface gravities comparable to GD659 and GD71, two of the cooler objects in our collection, the Wachter et al. method would fail to unearth companions later than $\sim$M9. The limit is even earlier for the hottest white 
dwarfs studied here (with similar or larger radii). However, a detailed examination of Figures 1a-d, top and bottom, reveals no convincing evidence for the presence of a spatially unresolved cool companion to any of the eight targets of the present study. Therefore, following the method outlined in the previous section, we have placed the approximate limits  given in Table~\ref{tab2} on the spectral type of a putative companion to each white dwarf. Subsequently, we have used  these limits to constrain the mass of each putative companion.

\subsection{Limiting masses to putative late-type companions}

Although the cooling age of each white dwarf can be estimated from theoretical evolutionary models 
(e.g Wood 1995), we don't know with any certainty the mass and hence lifetime of their 
progenitors. Therefore, we are unwilling to assign an age to the putative associate 
of any of our white dwarfs (ie. progenitor lifetime + wd cooling time). As the effective temperatures of 
very-low-mass stars and substellar objects remain a function of both mass and age at times
of the order gigayears, when using our spectral-type limits to constrain masses, we instead assume a 
range of ages broadly encompassing likely values. We have used the polynomial fit detailed in Table 4 of Golimowski et al. (2004) to 
assign approximate effective temperatures to the spectral type limits shown in Table~\ref{tab2}. Subsequently, we 
refer to the low mass stellar/substellar evolutionary models for solar metallicity of Baraffe et al. 
(2003), using cubic splines to interpolate between their points, to estimate corresponding masses at ages 1Gyr, 5Gyrs 
and 10Gyrs as shown in Table~\ref{tab2}. Clearly, even for an age comparable to that of the Galactic 
disk our results argue strongly against the presence of a companion with M$\simgreat0.082$M$_{\odot}$ to any 
of the degenerates in our collection. Indeed, population synthesis models indicate that most of 
these white dwarfs are likely to be the progeny of local disk F stars (e.g. Aznor Cuadrado et al. 2004, Schroder, Pauli \& Napiwotzki 2004). As the 
lifetime of a F star is $\sim5$Gyrs (e.g de Loore \& Doom 1992), and the time it takes a DA white 
dwarf of canonical mass (0.6M$_{\odot}$) to cool to 12000K is $\sim0.5$Gyrs years, the majority of 
these objects likely formed less than 6Gyrs ago. 

\subsection{Overall relevance of results and significance to individual objects}

Considering the typical distances of the white dwarfs in this study ($\sim60$pc), the atmospheric conditions at the 
time the data were acquired and the width of the instrument slit, the spectroscopic nature of our observations renders 
our study most sensitive to companions at a~$\simless$60AU from our degenerate primaries. It is probable that the 
manner in which we have selected our targets, as described in Section 1, has biased our work against systems 
containing cool dwarfs earlier than mid-M spectral type. The presence of such stars would most likely have been 
detected as excess red continuum in the 
optical spectrum (e.g. Vennes \& Thorstensen 1994) of these relatively well studied white dwarfs. Hence this work
 is, in effect, sensitive to secondaries of mid-M type and later, which in terms of mass is 
M$\approx0.15-0.08$M$_{\odot}$ (Kirkpatrick, Henry \& McCarthy 1991; Baraffe et al. 2003). The spectral types of
the 10 cool companions unearthed from the  EUV selected sample of white dwarfs were estimated to range from 
$\sim$M3.5-M6.5 (M$\approx0.3-0.1$M$_{\odot}$; Green, Ali \& Napiwotzki 2000; Kirkpatrick, Henry \& McCarthy 1991).
Neglecting that poorer seeing and a slightly more distant white dwarf sample probably resulted in their photometric 
study having some additional sensitivity to companions at wider separations, if we assume a mass function for the
secondary stars of dN/dM$\propto$M$^{-1}$ (e.g. Reid \& Gizis 1997), on the basis of the Green, Ali \& Napiwotzki 
result, we can estimate, albeit rather crudely, that we might have expected to detect late-type companions to 
$\sim$ 1 in 6 of the white dwarfs in the present collection.

Although the small number of white dwarfs in our study has likely contributed significantly to our failure to 
detect a late-type secondary, it has become apparent from the results of detailed radial velocity surveys that there 
is a deficiency of very-low-mass stellar and substellar companions to F,G and K type main sequence stars at 
separations less than $\sim5$\,AU (q(M$_{2}$/M$_{1}) \simless0.2$; e.g Halbwachs et al. 2003, Marcy \& Butler 2000).
Recent coronographic near-IR imaging of more than 250 nearby, young ($\simless300$Myrs) G,K and M
dwarf main sequence stars indicates that this ``brown dwarf desert'', as it has become known, extends out to 
$\sim1000$\,AU (McCarthy \& Zuckerman 2004). While it could be argued that very-low-mass stellar and substellar 
companions lying within a few AU of a white dwarf's progenitor would be obliterated during a common envelope phase 
of post main-sequence evolution, binary evolution models indicate that unless the transfer of energy from the 
companions orbit to the stellar envelope is extremely inefficient, a significant proportion of very-low-mass 
secondaries will survive this phase (Politano 2004). Furthermore, our study is sensitive to companions lying 
well beyond the typical radius of the envelope of an AGB star ($\sim1-2$AU, Schenker, 2004, private comm.), even 
allowing for the change in separation which likely occurs as mass is lost from the white dwarf progenitor. It 
therefore seems plausible that our search overlaps with the top end of the the brown dwarf desert in terms of mass 
and that this may also have had a bearing on the outcome of our search. We note that Farihi et al. (2003), who
are conducting a proper motion survey of white dwarfs based on IR imaging, find very few late-M type and cooler 
companions, albeit at relatively wide separation (hundreds of AU). Despite our rather small and disparate collection
of targets, it seems fair to say that the present results don't contradict previous findings regarding very-low-mass 
companions to main sequence stars and white dwarfs ie. we have not unearthed a previously unrecognised population 
of late-M and L type companions to these DA white dwarfs.
 
These results also have implications for our objects at an individual level, since several are known to exhibit 
anomolies in their photospheric abundance patterns (e.g. GD659) or hydrogen line profiles (e.g. WD2359-438):

{\bf GD659:} This white dwarf displays an EUV energy distribution consistent with a near pure-H photosphere 
(e.g Barstow et al. 1997), despite the presence in the atmosphere of C, N and Si as revealed by STIS and IUE  
spectroscopy (e.g. Barstow et al. 2003a). However, detailed inspection of the NV line profiles suggests that 
this element is not homogeneosly distributed in depth, with the bulk confined to higher, lower pressure layers 
of the photosphere (Barstow et al. 2003a). According to these authors this high degree of stratification is 
not predicted by radiative levitation theory. Using our near-IR spectroscopy we are able to rule out that these
abundance anomalies arise from ongoing accretion of material from an unseen companion of spectral type L6 or earlier.
Refering to the 10Gyr model of Baraffe et al. (2003) we find that this amounts to the exclusion of a
(spatially unresolved) stellar companion to GD659.

{\bf GD50:} The EUVE spectrum of this white dwarf reveals the presence of photospheric helium at an abundance of 
log(He/H)$\approx-3.6$. This cannot be explained in terms of radiative levitation theory, which predicts 
log(He/H)$\simless-8.0$ for an object of GD50's effective temperature and surface gravity (Vennes et al. 1996). 
An alternative 
source for this helium is material being accreted into the atmosphere from an unseen low mass companion. However, 
Vennes et al. set a limit of M7-M8 on the spectral type of any such object on the basis of I band photometry. 
Using our near-IR spectroscopy we are now able to push this limit to spectral type T5, which corresponds to a 
mass of $\sim69$M$_{\rm Jup}$ at an age of 10Gyrs. Thus it seems increasingly likely that presence of helium in 
GD50 is in some way related to its unusually large mass (M$=1.23\pm0.05$M$_{\odot}$; Marsh et al. 1997), and its
possible formation via a stellar merger (e.g. Guerrero, Garcia-Berro \& Isern 2004). 

{\bf RE J0457-280:} The IUE spectrum of this white dwarf reveals the presence of C,N and O in the photosphere. 
However, the steep drop in the EUV flux shortward of 250\AA\ indicate that Fe (and perhaps Ni too) is also likely 
present in significant quantities. Surprisingly, Barstow et al. (2003b) find a much larger discrepancy between the
 Lyman and Balmer derived effective temperature determinations, compared to that observed for other DAs of similar effective temperature and surface gravity. Despite the 2MASS H-K$_{\rm S}$ colour hinting at the presence of a cool companion this is not confirmed by
 our near-IR spectroscopy. Instead we rule out an unresolved associate of spectral type L3 or earlier, corresponding to M$\simgreat0.075$M$_{\odot}$ at 10Gyrs (Baraffe et al. 2003). 

{\bf GD71:} The EUV energy distribution of GD71 is consistent with a near pure-H atmosphere 
(e.g Barstow et al. 1997). The co-added IUE echelle spectrum provides no convincing evidence from the 
presence of photospheric metals (Holberg, Barstow \& Sion 1998). Furthermore, the radial velocity of this white 
dwarf shows no 
significant variability (Maxted, Marsh \& Moran 2000). Thus we have no compelling reason to suspect
the existence of a very-low-mass companion. However, on the basis of our near-IR spectroscopy we are now able to
exclude a (spatially unresolved) stellar companion to GD71.  

{\bf RE J2214-490:} Enhancements in the C,O, Fe and Ni abundances in this star with respect to G191-B2B (Barstow et al. 2003a), once again raise the possibility of ongoing accretion from an unseen companion. However, from the near-IR spectroscopy we can rule out an unresolved companion of spectral type M9 or earlier, corresponding to M$\simgreat0.082$M$_{\odot}$ at 10Gyrs (Baraffe et al. 2003). Alternatively, increased equilibrium abundances of these elements may simply result from the stars slightly larger effective temperature and marginally lower surface gravity in comparison with G191-B2B (e.g Schuh et al. 2002). 

{\bf WD2359$-$434:} This white dwarf is relatively cool ($T_{\rm eff}=8660$\,K) and has a high
 mass ($0.95M_\odot$; Aznar
Cuadrado et al.\ 2004). Koester et al.\ (1998) reported a very shallow
and narrow H$\alpha$ profile and speculated that a magnetic field
might be responsible. Aznar Cuadrado et al.\ (2004) indeed detected a
weak magnetic field of 3.1\,kG during their spectropolarimetric survey
of bright white dwarfs. However, in order to explain the observed
profile of the H$\alpha$ core, one has to invoke a magnetic field with
a maximum field strength of $>$50\,kG.  Although it cannot be
completely ruled out that both observations can be explained by a
peculiar and complex magnetic field structure, other explanations have
to be explored as well. 

Due to the low temperature and the small radius of WD\,2359$-$434 we
can set quite stringent upper limits on a cool companion. We derived a
limiting spectral type of T8 corresponding to 0.05\,$M_\odot$ for an
age of 10\,Gyr, i.e.\ a stellar companion can be ruled out.
Warm circumstellar dust causing an infrared excess was detected in the
ZZ\,Ceti star G29-38 (Tokunaga, Becklin \& Zuckerman 1990). No sign of such an infrared
excess is present in our spectrum of WD2359$-$434, which limits the
amount of circumstellar dust which could be present close to the white
dwarf.

\section{Conclusions}

Our detailed near-IR spectroscopic study of a collection of eight predominantly southern hemisphere DA white
dwarfs has failed to reveal the presence of any late-type companions. Instead, we have placed approximate 
limits on the spectral-types of putative companions. These constraints allow us to rule out spatially 
unresolved low-mass stellar associates to GD659,GD50,GD71 and WD2359-434 and companions with 
M$\simgreat0.082$M$_{\odot}$ to any of the remaining stars in the collection. These results argue against 
ongoing accretion of material from low mass companions as the source of the abundance anomalies seen in a 
number of these stars. Furthermore, they can be viewed as consistent with the previously reported drop below 
q$\sim0.1-0.2$ in the mass ratio distribution of binaries with main sequence F,G,K and M type primaries as determined 
from detailed radial velocity studies.
  
\section*{Acknowledgments}
PDD and AJL are sponsored by PPARC in the form of a postdoctoral grant and a studentship respectively. 
MRB and RN acknowledge the support of PPARC in the form of advanced fellowships.
JBH wishes to acknowlege support from NASA Grant No. NNG04GD87G. This publication makes use of data 
products from the Two Micron All Sky Survey, which is a joint project of the University of Massachusetts
and the Infrared Processing and Analysis Center/California Institute of Technology, funded by the National
Aeronautics and Space Administration and the National Science Foundation.

\appendix

\bsp

\label{lastpage}

\end{document}